\documentclass{sig-alternate-05-2015}
\begin{document}

\setcopyright{acmcopyright}

\doi{http://dx.doi.org/xx.xxxx/xxxxxxx.xxxxxxx}

\isbn{978-1-4503-4486-9/17/04}


\acmPrice{\$15.00}

%
\conferenceinfo{SAC'17,}{ April 3-7, 2017, Marrakesh, Morocco}
\CopyrightYear{2017} 

\title{MOMOS-MT: Mobile Monophonic System \\ for Music Transcription}
\subtitle{Sheet Music Generation on Mobile Devices}

%
%
%
%
%

\numberofauthors{4} 
%
\author{
%
%
\alignauthor
Munir Makhmutov\\
      \affaddr{AI in Games Dev. Lab}\\
      \affaddr{Innopolis University, Russia}\\
      \email{m.makhmutov@innopolis.ru}
\alignauthor
Joseph Alexander Brown\\
      \affaddr{AI in Games Dev. Lab}\\
      \affaddr{Innopolis University, Russia}\\
      \email{j.brown@innopolis.ru}
\and  
\alignauthor Manuel Mazzara\\
      \affaddr{Service Science and Engineering Lab}\\
      \affaddr{Innopolis University, Russia}\\
      \email{m.mazzara@innopolis.ru}
\alignauthor Leonard Johard\\
      \affaddr{Service Science and Engineering Lab}\\
      \affaddr{Innopolis University, Russia}\\
      \email{l.johard@innopolis.ru}
}
\date{30 July 1999}

\maketitle
\begin{abstract}
Music holds a significant cultural role in social identity and in the encouragement of socialization. Technology, by the destruction of physical and cultural distance, has lead to many changes in musical themes and the complete loss of forms. Yet, it also allows for the preservation and distribution of music from societies without a history of written sheet music. This paper presents early work on a tool for musicians and ethnomusicologists to transcribe sheet music from monophonic voiced pieces for preservation and distribution. Using FFT, the system detects the pitch frequencies, also other methods detect note durations tempo, time signatures and generates sheet music. The final system is able to be used in mobile platforms allowing the user to take recordings and produce sheet music \emph{in situ} to a performance.
\end{abstract}

%
%

\ccsdesc[500]{Applied computing~Sound and music computing}
\ccsdesc[500]{Information systems~Music retrieval}
\ccsdesc[300]{Information systems~Open source software}
%
%

%
%
\printccsdesc


\keywords{Monophonic Music Transcription; Music Analysis; Pitch Frequency Detection; Pitch Duration Detection; Tempo Detection; Time Signature Detection; Ethnomusicology}

\section{Introduction}
The complexity and variety of musical genres is one of the major obstacles for the delivery of an effective commercial system for polyphonic music transcript. However, obstacles for its adoption are mostly of cultural nature and not technical. Music  transcription is considered by musicians and music teachers a good pedagogical exercise, and there are prejudices in some communities about the liberalization of automatic tools. In some cases, though, written transcription is simply not possible, and this is mostly true for regional folk music that it is usually played and not written and the conservation process continued unchanged for centuries. In other cases, automatic music transcription would allow musicians to record the notes of an improvised performance that would otherwise get lost (this situation is typical of jazz) \cite{Benetos13}. Software technologies can be of paramount importance in this context to simplify the diffusion of local music artifacts, protect them, and allow musicians in other parts of the globe to access them via Internet in a way that would have been unthinkable only two decades ago.

Automatic music transcription can be traced back to the late 1970s when Martin Piszczalski and Bernard Galler aimed at creating ``a system that would automatically transcribe musical sounds into their equivalent music notation''. The objective of their project was to ``automatically detect the notes in a played melody and write out the note patterns just as a human transcriber would'' \cite{Piszczalski1979}. At the time, available technologies suggested the use of recorded tapes, and computers would analyze such tapes with music recorded from musical instruments. The researchers understood that their technique could have benefited several fields such as music education and music publishing. Their contribution, although among the first fully comprehensive from a methodological viewpoint, was not built in the vacuum. The idea of using a computer program to automatically convert sounds into written music appeared together with the widespread use of computers, although systematic methodologies took longer to appear \cite{Askenfelt1976}. Since the first studies on music transcription, one of the major challenges was the development of polyphonic systems, which appeared significantly more problematic. The seminal work by Moorer \cite{Moorer75} was then continued by researchers at Stanford in the beginning of the 1980s \cite{Chafe82,Chafe86}, still the systems did not allow the transcription of more than two polyphonic voices.

Despite the analogies with speech recognition, the commercial and academic interest for music transcription has been more moderate and has been mostly pushed by interdisciplinary research with broader goals than transcription itself. The automatic transcription of polyphonic music has been the subject of increasing interest during the last thirty years. Ideas that had originated in the 1970s encouraged research on the topics for the following decades and influenced different research fields, in particular coping with the limitation of monophonic transcription systems. Work has been done in multidisciplinary contexts too, for example in the crossover between robotics, AI and music transcription. One of the first projects allowing the transcription of more than two polyphonic voices started in the late 1980s at Osaka University in Japan and was aimed at extracting feelings from musical signals to form the software core of a robotic system that could respond to music as human listeners \cite{Katayose89}.


In this paper, we tackle the problem of automatic transcription of monophonic music via mobile device recording, and we present the tool MOMOS-MT (Mobile Monophonic System for Music Transcription).

Mobile devices are the major enabler for sheet music creation \emph{in situ} to a performance. The authors believe this aspect represents the major driver of innovation and foresee a potential for commercialization to support the progress of ethnomusicology.

This paper is organized as follows. Section~\ref{problemsolution} describes in detail the problem description and the known techniques for pitch and tempo recognition. Section~\ref{sys} is dedicated to the description of transcription system on which this paper is built. Section~\ref{conclusions} reports the overall results.

\section{Problem Description}\label{problemsolution}

Methods and techniques for music transcription, in particular monophonic, have been investigated over four decades. In this section, we will present the literature in some detail for what concerns pitch (frequency and duration), tempo, and time signature detection.

\subsection{Pitch Recognition Methods}\label{PitchRecognitionMethods}
There are a lot of methods used for pitch detection, but not all of them give good results \cite{cook94, fucks62}. Some papers tell that autocorrelation can be used for monophonic transcription \cite{brown91, giuliano00, juan00}. Some papers recommend to use Fourier transform for music transcription: for single notes and for chords. Statistics of these two methods usage show that autocorrelation has worse accuracy of pitch recognition than Fourier transform. Chord recognition is more difficult task, it also can be called polyphonic music transcription \cite{kappen03, bello03}.

\begin{figure}
	\centering
    \includegraphics[width=.5\textwidth]{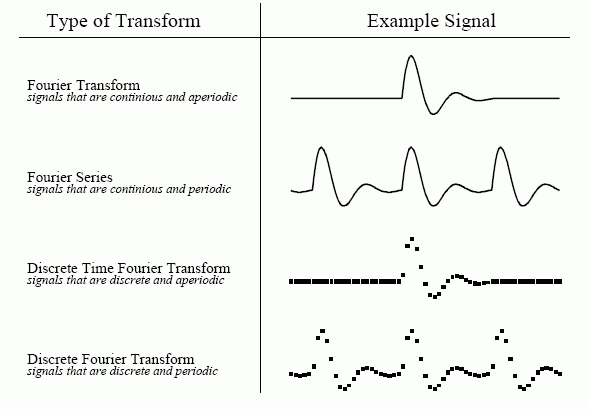}
    \caption{Fourier transform types \cite{smith99}}
    \label{fig:fourier}
\end{figure}

The Fourier transform decomposes a function of time (\emph{e.g.} a signal) into the frequencies that make it up. This process is similar to how a musical chord can be expressed as the amplitude or loudness of its constituent notes. The Fourier transform of a function of time itself is a complex-valued function of frequency, whose absolute value represents the amount of that frequency present in the original function, and whose complex argument is the phase offset of the basic sinusoid in that frequency. 
The term Fourier transform refers to both the frequency domain representation and the mathematical operation that associates the frequency domain representation to a function of time. The Fourier transform is not limited to functions of time, but in order to have a unified language, the domain of the original function is commonly referred to as the time domain. For many functions of practical interest, one can define an operation that reverses this: the inverse Fourier transformation, also called Fourier synthesis, of a frequency domain representation combines the contributions of all the different frequencies to recover the original function of time. There are four types of Fourier transform (Figure \ref{fig:fourier}).

A fast Fourier transform (FFT) is an algorithm to compute the discrete Fourier transform (DFT) and its inverse. Fourier analysis converts time, or space to frequency, or wavenumber and vice versa; an FFT rapidly computes such transformations by factorizing the DFT matrix into a product of sparse (mostly zero) factors. It is much faster than other Fourier transform types. 

In his PhD thesis, Sterian \cite{sterian99} has used Pielemeier's modal transform \cite{Pielemeier96} as a front end, the peaks of which were tracked through frames by Kalman filtering for polyphonic transcription \cite{kalman1960}. Reasonable success was found with up to four part brass ensembles (79.5\% accuracy was given in four-note polyphony).

Walmsley used Markov Chain Monte Carlo (MCMC) methods  to attempt the transcription task \cite{walmsley00}. The number of notes and harmonics were unknown. This produced reasonable results, but with the inclusion of global parameters controlling the evolution of data over multiple frames, performance was improved. Godsill \& Davy \cite{godsill02} continued this research and improved the model to take account of amplitude evolution within a single frame and also inharmonicity \cite{brown94}. Their simulations are computationally intensive but produce accurate results for traditionally difficult problems (\emph{e.g.} fifths).

Recently attention has been directed to the transcription of non-Western music, which has underlined both the capability and limitations of current transcription systems \cite{benetos2015automatic} applied the polyphonic transcription techniques developed in  \cite{benetos2012shift} in order to transcribe traditional Turkish microtonal music. Cited accuracies are 56.7 \% (F-measure). The same technique was applied by \cite{abdallah2015automatic} to the the British Library World \& Traditional Music Collections, but unfortunately accuracy estimate are not available. Another recent approach by \cite{holzapfel2016sousta} reached 57.9 \% F-measure on Greek dances.

\subsection{Pitch Frequency Precision}\label{pitchPrecision}
After recognition of note frequencies (no matter which recognition method was used), they should be changed. There are two reasons for changing of note frequencies:
\begin{enumerate}
  \item musical instruments cannot be ideally tuned
  \item even for well-tuned instrument pitch recognition methods cannot give ideal results
\end{enumerate}
As previously mentioned, MIDI file can contain a sequence of notes that has one of 128 basic meanings of frequencies. As such, it is necessary to compare each frequency with these 128 basic frequencies and find the nearest. Of course, it is naive to compare each of the 128 pitches; it is better to use one of the optimization methods that will get the necessary pitch in a small amount of iterations. 
\par The Dichotomy method can be used for this task because it can reduce the amount of iterations with a small amount of equations \cite{wilde}. It is a method for numerically solving equations with a single unknown. Consider the equation $$f(x)=0$$
with a continuous function 
\begin{math}
f
\end{math}
on the interval
\begin{math}
[a,b]
\end{math}
which takes values of different signs at the end points of the interval and which has a single root
\begin{math}
x^*
\end{math}
within
\begin{math}
[a,b]
\end{math}.
To find
\begin{math}
x^*
\end{math}
approximately, one divides
\begin{math}
[a,b]
\end{math}
into halves and calculates the value of
\begin{math}
f(x_1)=0
\end{math}
at the midpoint
\begin{math}
x_1=(a+b)/2
\end{math}.
If
\begin{math}
x_1\ne0
\end{math},
one takes the two intervals
\begin{math}
[a,x_1]
\end{math}
and
\begin{math}
[x_1,b]
\end{math}
and from them selects for the next dichotomy the one at the end points of which the values of the function differ in sign. This continued division into halves gives a sequence
\begin{math}
x_1,x_2,\dots,x_n,
\end{math}
which converges to the root
\begin{math}
x^*
\end{math}
with the rate of a geometrical progression:
\begin{center}
\begin{math}
|x_n-x^*|\le \frac{b-a}{2^n}, n=1,2,\dots (1)
\end{math}
\end{center}
\noindent where the bound (1) cannot be improved upon in this class of functions. If
\begin{math}
f
\end{math}
has more than one root in
\begin{math}
[a,b]
\end{math},
the sequence will converge to one of them.

A method for minimizing a function of one variable. One has to find the minimum
\begin{center}
\begin{math}
f^*=\min_{x \in [a, b]}  f(x)
\end{math}
\end{center}
of a unimodal function
\begin{math}
f
\end{math}
on an interval
\begin{math}
[a,b]
\end{math}
and to determine the point
\begin{math}
x^*
\end{math}
at which it is attained. For this, one divides
\begin{math}
[a,b]
\end{math}
into halves and near the middle
\begin{math}
\overline{x_1}=(x+b)/2
\end{math}
calculates the values of
\begin{math}
f
\end{math}
at the two points
\begin{math}
x_1=\overline{x_1}-\varepsilon/2
\end{math}
and
\begin{math}
x_2=\overline{x_1}+\varepsilon/2
\end{math},
where the number
\begin{math}
\varepsilon>0
\end{math}
is a parameter of the method and is sufficiently small. Then, the values
\begin{math}
f(x_1)
\end{math}
and
\begin{math}
f(x_2)
\end{math}
are compared, and on the basis that
\begin{math}
f
\end{math}
is unimodal one selects from the two intervals 
\begin{math}
[a,x_2]
\end{math}
and
\begin{math}
[x_1,b]
\end{math}
the one that certainly contains
\begin{math}
x^*
\end{math}.
For example, if 
\begin{math}
f(x_1)\ge f(x_1)
\end{math},
this will be 
\begin{math}
[a,x_2]
\end{math},
otherwise
\begin{math}
[x_1,b]
\end{math}.
The interval is again divided into halves, and near the middle
\begin{math}
\overline{x_2}
\end{math}
one takes two points
\begin{math}
x_1=\overline{x_2}-\varepsilon/2
\end{math}
and
\begin{math}
x_2=\overline{x_2}+\varepsilon/2
\end{math},
compares the values of the function, etc. As a result, one obtains a sequence of midpoints
\begin{math}
$\{$\overline{x_n}$\}$
\end{math},
for which
\begin{center}
\begin{math}
|x_n-x^*|\le \frac{b-a-\varepsilon}{2^n}+\frac{\varepsilon}{2}, n=1,2,..., (2)
\end{math}
\end{center}
As an approximation to
\begin{math}
f^*
\end{math}
the value
\begin{math}
f(\overline{x_n})
\end{math}
for sufficiently large
\begin{math}
n
\end{math} is taken.

The name is given to the method because at each step in this algorithm the segment containing the minimum becomes approximately half the length. The dichotomy method is not the best in the class of unimodal functions. 
\par According to the number of MIDI levels (128), this algorithm will find the necessary frequency in seven-eight iterations. It is much better than 128 iterations. Other optimization methods also can be used. The disadvantage of other methods is that they are more difficult and will increase the amount of calculations, even the method of Golden section search, a type of modified Dichotomy method.

\subsection{Tempo Detection}\label{tempo}
Tempo detection is one of the first tasks in music transcription. Mistake in tempo detection can lead to a high error probability of music transcription, which justifies our thoroughness in selecting methods for tempo recognition. One of the methods is Monte Carlo methods for tempo tracking \cite{cemgil03}. Further, there are several open source libraries for different languages that can be used for this task. Tempo is not always constant in songs: it can vary throughout the song. The issue of tempo tracking is that it can be changed slowly and immediately. For this reason, tempo tracking can become a difficult task, especially when time signature changes as well. Most popular music is written in constant tempo; therefore, it is assumed in our method that songs have a constant tempo. 

\subsection{Time Signature Detection}\label{timeSignature}
As mentioned earlier, time signature can vary too, and it also can be constant during the song. Although non-constant time signature is not common for modern music, it is used in some songs. Detection of changing time signatures is not a trivial task. There are common signatures, such as quadruple (4/4) and triple (3/4) time. Most songs are written in this time and it is constant. However, some songs are written in less frequent time signatures, such as 5/4, 7/4, 9/4, etc. are more difficult to detect. Especially, when a time signature changes mid-song.


\subsection{Pitch Duration Detection}\label{pitchDuration}
Each note has its duration, and there are several approaches used for detecting pitch duration. One of these approaches provides measuring of loudness and energy for each small part of song. So, higher loudness can mean that this part of song is represented by a new pitch. Of course, this method will work only if the previous note is considered. It is necessary because neighboring chunks of the song with the same meaning of pitch can be parts of one note or it can be two notes played consequently. If the second chunk has the same pitch and lower or same loudness, then it is most likely the same note. If the second chunk has the same pitch and higher loudness, then it is most likely two subsequent notes with the same pitch. Additional analysis is needed to make a decision about it.

\section{Transcription System}\label{sys}

\begin{figure}[h]
	\centering
    \includegraphics[width=.45\textwidth]{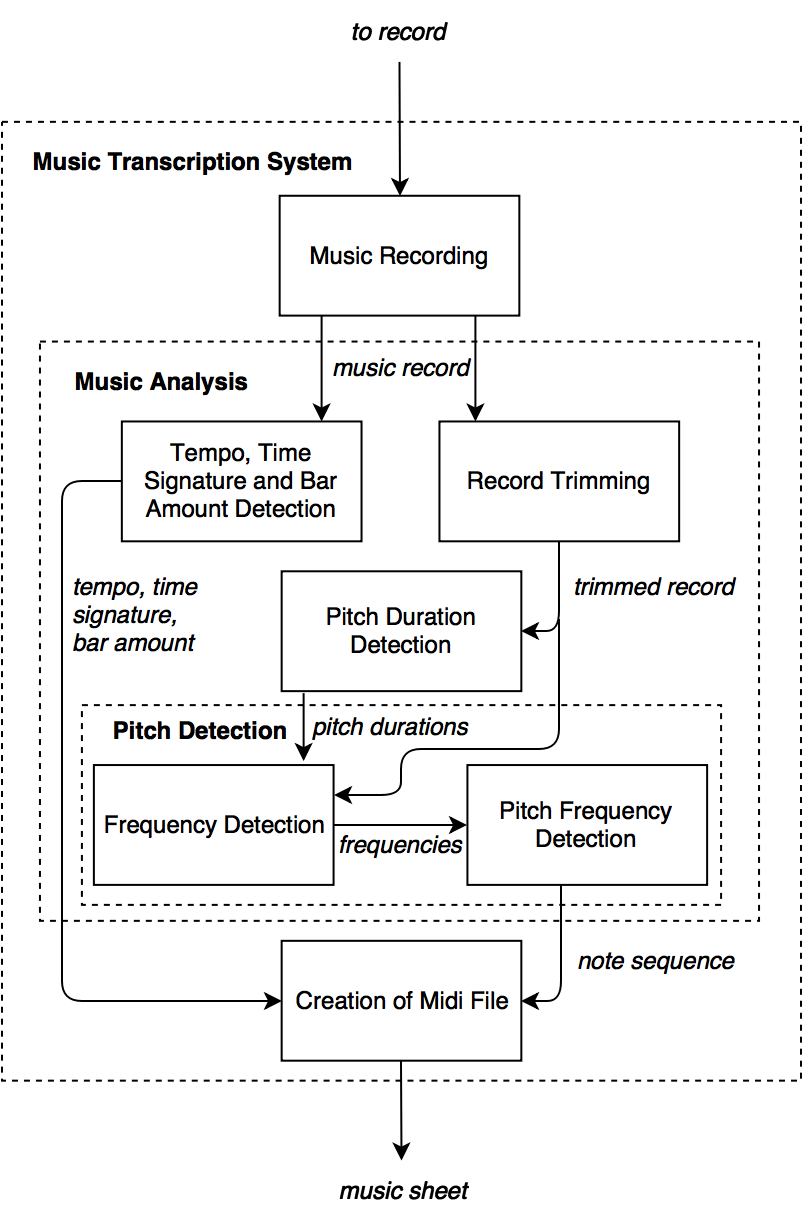}
	\caption{Music Transcription Process}
	\label{fig:mtp}
\end{figure}

A transcription system prototype has been developed as a part of the research. It works mostly as a \emph{proof of concept} in order to verify the potential for practical applications. Figure~\ref{fig:mtp} describes the music transcription process as realized by the tool.  

The music that has been recorded acts as input for the Music Analysis block. Here the recorded music is trimmed in order to detect tempo, time signature, bar amount, notes and their duration, which are all concepts explained in Section~\ref{problemsolution}. Finally, and accordingly to the processed data, the tool generates a MIDI containing the information for creating sheet music. This sheet music represents the final output of the process. The prototype source code is available on GitHub at the URL: \emph{hidden for peer review}.
At the same URL, a set of recordings used to test the system is also available.

The key feature of the developed system is the ability to produce sheet music \emph{in situ} to a performance. Using mobile devices has not only enabled, but also propelled the innovation for this prototype. The development has been driven by the definition of three principal use cases derived by the elicitation of functional requirements: music recording, music transcription, and transcription sharing. The use cases also describe the lifecycle of the mobile application and the the main steps of function.

\subsection{Requirements}\label{FR}

This subsection enlists the necessary requirements that are required for the mobile monophonic music transcription system development. These requirements have been elicited from interactions with musicians, potential users, and specialists of Music Informatics from universities in Europe (\emph{hidden for peer review}).
Requirements from 1 to 5 have been fully implemented in the current prototype while the option of sharing created files via email is still under development and left as future work. At the same time, other sharing options are being considered under the wave of increasing use of social media by music experts.

\begin{center}
\begin{tabular}{ | p{2.9cm} | p{4.8 cm} | } 
 \hline
 \textbf{Requirement ID} & FR-01 \\
 \hline
 \textbf{Title} & Application should be available for mobile phones and tablets \\
 \hline
 \textbf{Description} & Application should be available on Android devices \\ 
 \hline
 \textbf{Priority} & Mandatory (1) \\ 
 \hline
 \textbf{Risk} & Critical (C) \\ 
 \hline
\end{tabular}
\end{center}

\begin{center}
\begin{tabular}{ | p{2.9cm} | p{4.8cm} | } 
 \hline
 \textbf{Requirement ID} & FR-02 \\
 \hline
 \textbf{Title} & Application should allow user to record the music \\
 \hline
 \textbf{Description} & Application has to have an opportunity to record the music on given device via internal microphone \\
 \hline
 \textbf{Priority} & Mandatory (1) \\ 
 \hline
 \textbf{Risk} & Critical (C) \\ 
 \hline
\end{tabular}
\end{center}

\begin{center}
\begin{tabular}{ | p{2.9cm} | p{4.8cm} | } 
 \hline
 \textbf{Requirement ID} & FR-03 \\
 \hline
 \textbf{Title} & Application should allow user to transcribe monophonic music played by only one instrument \\
 \hline
 \textbf{Description} & Application should allow user to transcribe monophonic music with accuracy not less than 90$\%$. Monophonic music provides single instrument that plays not more than one note per each moment \\ 
 \hline
 \textbf{Priority} & Mandatory (1) \\ 
 \hline
 \textbf{Risk} & Critical (C) \\ 
 \hline
\end{tabular}
\end{center}

\begin{center}
\begin{tabular}{ | p{2.9cm} | p{4.8cm} | } 
 \hline
 \textbf{Requirement ID} & FR-04 \\
 \hline
 \textbf{Title} & Application should allow user to save file with music transcripts. There are several file types that can be used: MusicXML, ASCII, MIDI, gpx \\
 \hline
 \textbf{Description} & Application should save extracted music into file with popular extension that can be read and changed on computer \\ 
 \hline
 \textbf{Priority} & Mandatory (1) \\ 
 \hline
 \textbf{Risk} & Critical (C) \\ 
 \hline
\end{tabular}
\end{center}

\begin{center}
\begin{tabular}{ | p{2.9cm} | p{4.8cm} | } 
 \hline
 \textbf{Requirement ID} & FR-05 \\
 \hline
 \textbf{Title} & Application should eliminate excess noise from the record \\
 \hline
 \textbf{Description} & Application should eliminate excess noise in the record, which can lead to wrong music transcription \\ 
 \hline
 \textbf{Priority} & Nice to have (2) \\ 
 \hline
 \textbf{Risk} & Medium (M) \\ 
 \hline
\end{tabular}
\end{center}

\begin{center}
\begin{tabular}{ | p{2.9cm} | p{4.8cm} | } 
 \hline
 \textbf{Requirement ID} & FR-06 \\
 \hline
 \textbf{Title} & Application should allow user to send created file via email \\
 \hline
 \textbf{Description} & Application should have menu option that can help to send created file via email \\ 
 \hline
 \textbf{Priority} & Optional (3) \\ 
 \hline
 \textbf{Risk} & Low (L)  \\ 
 \hline
\end{tabular}
\end{center}

\subsection{Use Cases}\label{UC}

The use cases define the interaction between a human user (in this case) and the system, for achieving specific goals: \emph{recording} music, \emph{transcribe} a music file and \emph{share} a music file via email. The three cases cover part of the functional requirements that have been elicited from musicians and experts for the system design. Use Case 3 is not implemented in the current prototype.


\begin{center}
\begin{tabular}{ | p{2cm} | p{5.7cm} | } 
\hline
\textbf{Use case ID} & UC-01 \\
\hline
\textbf{Title} & Record music file \\
\hline
\textbf{Aim} & To make record with mobile/tablet \\ 
\hline
 \textbf{Reqs} & FR-1, FR-2, FR-5 \\ 
 \hline
 \textbf{Stakeholder interest} & Music recording for further transcription \\ 
  \hline
\textbf{Precondition} & 
1. Application is launched 
\par2. Enough free memory on device for record \\
  \hline
  \textbf{Main success scenario} &  
1. User chooses recording option
\par2. User records music via internal device microphone
\par3. Application eliminates excess noise from the record
\par4. User assigns name to the record
\par5. User saves record \\
  \hline
  \textbf{Post condition} & Record is saved on device \\
  \hline
\end{tabular}
\end{center}


\begin{center}
\begin{tabular}{ | p{2cm} | p{5.7cm} | } 
\hline
 \textbf{Use case ID} & UC-02 \\
 \hline
 \textbf{Title} & Transcribe music file \\
 \hline
 \textbf{Aim} & To transcribe record made with mobile/tablet \\
 \hline
 \textbf{Reqs} & FR-3, FR-4 \\ 
 \hline
 \textbf{Stakeholder interest} & Music transcription \\ 
 \hline
 \textbf{Precondition} & 
1. Monophonic music file was recorded
\par 2. Enough free memory on device for file transcription \\
 \hline
 \textbf{Main success scenario} &  
1. User chooses transcription option
\par 2. User assigns name to the file with transcriptions
\par 3. User saves file with transcriptions \\
 \hline
 \textbf{Post condition} & Music transcription is done, file is saved on device \\
 \hline
\end{tabular}
\end{center}


\begin{center}
\begin{tabular}{ | p{2cm} | p{5.7cm} | } 
 \hline
 \textbf{Use case ID} & UC-03 \\
 \hline
 \textbf{Title} & Send transcribed music file via email \\
 \hline
 \textbf{Aim} & To send transcribed music file  via email  \\
 \hline
 \textbf{Reqs} & FR-6 \\ 
 \hline
 \textbf{Stakeholder interest} & Music transcription \\ 
 \hline
 \textbf{Precondition} & 
1. Music transcription file was created
\par 2. Internet connection exists \\
 \hline
 \textbf{Main success scenario} &  
  1. User chooses option of sending file via email
  \par 2. User enters email address
  \par 3. User sends file via email \\
 \hline
 \textbf{Post condition} & Music transcription is sent via email \\
 \hline
\end{tabular}
\end{center}


\subsection{Methods and Libraries}\label{Methods}
The transcription system was developed in Java by means of a number of libraries:
\begin{itemize}
\item \emph{jMusic} (http://explodingart.com/jmusic) has been used for reading files in WAVEform audio format.
\item \emph{jEN} (https://github.com/echonest/jEN), an open source library, has been used for tempo, bar amount, time signature, and tempo tracking.
\item \emph{musicg} (https://code.google.com/p/musicg) has been used for record trimming (it was necessary to trim empty space at the beginning and at the end of records).
\item FFT has been implemented by the open source library \emph{Jtransforms} in combination with Dichotomy method have been used for notes' frequencies detection. Pitch duration detection is done by modified method of windowing\footnote{https://sites.google.com/site/piotrwendykier/software/jtransforms}.
\item The open source library \emph{jFugue} (http://www.jfugue.org) has been used to generate MIDI-files based on results input analysis.
\end{itemize}

\subsection{Running the System}\label{example}
We will show here how the system runs, \emph{i.e.} how a piece of music recording leads finally to the generation of a music sheet. As examples, we consider two melodies that were used as inputs for MOMOS-MT. 

The first melody (https://goo.gl/a6juhn) was played on a piano, while the second (https://goo.gl/pP4n2i) was played on a guitar. Both were recorded with a mobile phone recorder and outputs resulted in MIDI-files. The music was not transcribed taking into account the specific instruments (Output sample 1: https://goo.gl/n0ZRZV, Output sample 2: https://goo.gl/lTTeyV). The transcription system does not indeed detect instruments, which is why both melodies were transcribed into piano batches. The instrument for the second output was manually changed to guitar in order to have an easier comparison. Post-transcription changes of the instruments do not alter pitches, time signatures, or tempo. The only difference is the sound of the instrument, or timbre. 

\begin{figure}[h]
	\centering
    \includegraphics[width=.48\textwidth]{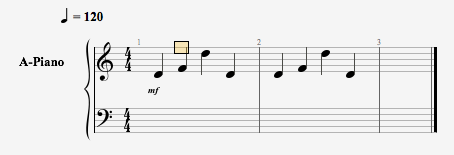}
    \caption{Output for the first record}
    \label{fig:output1}
\end{figure}

\begin{figure}
	\centering
    \includegraphics[width=.48\textwidth]{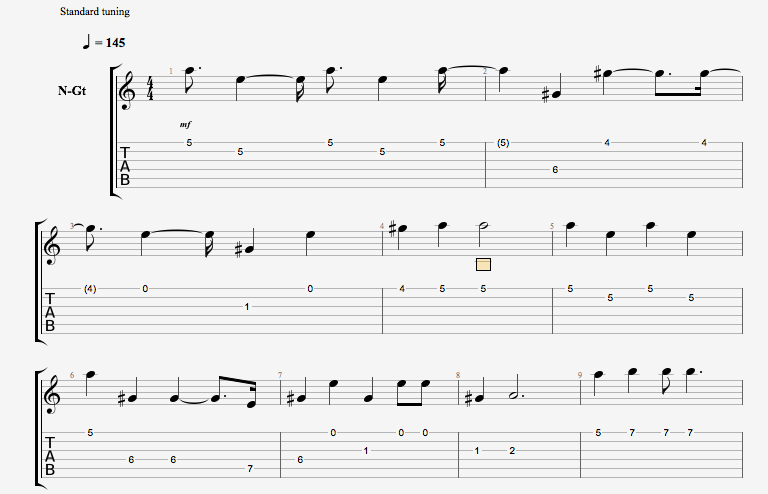}
    \caption{Output for the second record}
    \label{fig:output2}
\end{figure}

Sheet music for the output and part of the second are shown in Figure~\ref{fig:output1} and Figure~\ref{fig:output2}, respectively.

\section{Conclusions}\label{conclusions}
In this paper, the problem of automatic transcription of monophonic music with the usage of mobile device recorder has been considered, and the tool MOMOS-MT has been presented. The use of mobile devices is the key enabler of sheet music creation \emph{in situ} to a performance and represent the major driver of innovation and potential for commercialization of this prototype in the context of ethnomusicology. 
\par Fast Fourier transform was proposed to detect pitch frequencies. We have discussed and studied the musical theory in order to examine the system's requirements . We have found features, with which music can be described. The goal of the project was to create a mobile application for monophonic music transcription, which was achieved. 
\par The system outputs may differ from the inputs. The cause of this discrepancy is currently under evaluation, and a systematic asessment is left as future work. Anecdotal evidence shows these errors produce correct notes, but with a one-octave shift without violating the melody, as may be the case with other potental error types.
\par This work has described a transcription method that can be easily extended in the future. The next step for this system is polyphonic music transcription. Also, this system can be extended in the field of musical instrument recognition \cite{eronen00, BuccoliZSAS15} and the detection of different playing techniques on different instruments. For example, tapping, vibrato \cite{bendor00}, staccato, etc. Furthermore, dance music characteristics can be found, so the system will determine how popular transcribed music will be in disco clubs \cite{dixon03}. The direction of the sound also can be found, which can be useful for tracking systems \cite{davies01, martin99}. Marco Scirea \emph{et al.} \cite{scirea2016metacompose} created a tool for music generation. We plan future testing of MOMOS-MT to be performed taking as inputs the music generated by the tool. These future directions open new unexplored topics that will undoubtedly lead to new possibilities for all musicians of the world.







\bibliographystyle{abbrv}
\bibliography{bibliography}  
\end{document}